\documentclass[onecolumn,showpacs]{revtex4}
\usepackage{amssymb}
\usepackage{makeidx}
\usepackage{amsmath}
\usepackage{graphicx}
\usepackage{epstopdf}
\usepackage{dcolumn}
\usepackage{bm}

\input{tcilatex}
\begin{document}

\title{Microscopic nonlinear quantum theory of absorption of strong
electromagnetic radiation in doped graphene}
\author{A.K. Avetissian}
\author{A.G. Ghazaryan}
\email{amarkos@ysu.am}
\author{Kh.V. Sedrakian}
\author{B.R. Avchyan}
\affiliation{Centre of Strong Fields, Yerevan State University, 1 A. Manukian, Yerevan
0025, Armenia }
\date{\today }

\begin{abstract}
Microscopic quantum theory of nonlinear stimulated scattering of 2D massless
Dirac particles in doped graphene on Coulomb field of impurity ions at the
presence of an external strong coherent electromagnetic radiation is
developed. We consider high Fermi energies and low frequencies (actually
terahertz radiation) to exclude the valence electrons excitations. The
Liouville-von Neumann equation for the density matrix is solved
analytically, taking into account the interaction of electrons with the
scattering potential in the Born approximation. With the help of this
solution, the nonlinear inverse-bremsstrahlung absorption rate for a grand
canonical ensemble of 2D Dirac fermions is calculated. It is shown that one
can achieve the efficient absorption coefficient by this mechanism.
\end{abstract}

\maketitle



\section{Introduction}

Due to the known properties of graphene \cite{1,2,3,4,5,6} the coefficient
of interaction with the external electromagnetic (EM) radiation is very high
in comparison with other systems, which opens a significant field of
important applications of graphene in Nano-Opto-Electronics \cite%
{7,7a,7aa,7aaa,Mer2,Mer3,Mer4,Mer5,Mer6,Mer7,Xu2016}, as well as in
low-energy physics, condensed matter physics, and quantum electrodynamics
(QED).

High absorption coefficient \cite{trans} indicates that graphene strongly
interacts with light, and, it is very important that, because of gapless
structure of graphene such interaction can be efficiently realized with THz
devices as a multitron converter, as well as a protective material for
nanodevices \cite{18,19}. The strong absorption of electromagnetic (EM)
radiation in ultrasmall volumes (nanoscale) is a highly desirable property
for shielding materials used in nanoelectronics, the aerospace industry,
where strict requirements exist such as lightness and smallness or tightness.

Among the important processes induced by external radiation fields the
multiphoton stimulated bremsstrahlung (SB) is a basic mechanism of energy
exchange between the charged particles and plane monochromatic wave in
plasma-like media to provide the energy-momentum conservation law for real
absorption-emission processes that has been revealed immediately after the
invention of lasers \cite{12a}. What concerns the electrons elastic
scattering on impurity ions in graphene, there are many papers with
consideration of this basic scattering effect which have been described
mainly within the framework of perturbation theory by electrostatic
potential (see, e.g., \cite%
{BornEllastic,Chen2008,BornEl1,BornEl2,BornEl3,PhysRev2015,TanAdam2007,Katsnelson,Kaikai Xu17}%
). Regarding the SB process in graphene at moderate intensities of
stimulated radiation, in case of its linear absorption by electrons (or
holes), at the present time there are extensive investigations carried out
in the scope of the linear theory, see, e.g. \cite{7b,7b1,7b2,7b3,Zhu2014}.

Taking into account the above mentioned unusual high EM nonlinearity of
graphene, in this paper we have studied both analytically and numerically
the nonlinear absorption process of external EM radiation in doped graphene.
As a mechanism for the real absorption/emission of a plane-monochromatic
wave by the charged particles (or plasma-like medium), we have assumed SB
process of conductive electrons scattering on the charged impurities in
doped graphene. We developed microscopic quantum theory of graphene
nonlinear interaction with the coherent EM radiation of arbitrary intensity
and frequency. With the help of the solution of Liouville-von Neumann
equation for the density matrix, we calculated the nonlinear stimulated
scattering of 2D Dirac particles in graphene on the Coulomb field of
impurity ions at the presence of an external EM radiation field, taking into
account the interaction of electrons with the scattering potential in the
Born approximation. Here the selected frequency range of terahertz radiation
excludes the valence electrons excitations at high Fermi energies.

In Sec. I the relativistic quantum dynamics of SB of conductive electrons in
graphene is presented with analytical results for density matrix and
inverse-bremsstrahlung absorption rate. In Sec. II the analytic formulas in
case of screened Coulomb field of an impurity ion are considered
numerically. Finally, conclusions are given in Sec. III.

\section{Basic theory}

Interaction of a free electron with the EM wave is described by the
dimensionless relativistic invariant parameter of intensity $\xi
=eE_{0}\lambda /(2\pi mc^{2})$\ \cite{7}, which represents the wave electric
field (with amplitude $E_{0}$)\ work on a wavelength $\lambda $\ in the
units of electron rest energy. Apart, for THz photons $\hbar \omega \sim
0.01 $\ $\mathrm{eV}$, multiphoton effects take place at $\xi \sim 1$\ that
corresponds to intensities $I_{\xi }\sim 10^{14}$\ $\mathrm{Wcm}^{-2}$,
while the massless electron-wave interaction in graphene is characterized by
the dimensionless parameter $\chi =ev_{F}E_{0}/\left( \hbar \omega
^{2}\right) $\ \cite{7a}, which represents the work of the wave electric
field on a period $1/\omega $\ in the units of photon energy $\hbar \omega $%
. Depending on the value of this parameter $\chi $, three regimes of the
wave-particle interaction may be established: $\chi \ll 1$\ -- that
corresponds to one-photon interaction regime \cite{9,10,11}, $\chi \gg 1$ --
which is the static field limit of superpower fields in QED or Schwinger
regime \cite{12}, and $\chi \geqslant 1$\ -- is the multiphoton interaction
regime \cite{7a} with the corresponding intensity $I_{\chi }=\chi ^{2}\times
3.07\times 10^{11}$\ $\mathrm{Wcm}^{-2}\mathrm{[\hbar \omega /eV]}^{4}$.
Comparison of this intensity threshold with the analogous one for the free
electrons or with the situation in common atoms shows the essential
difference between the values of these thresholds: $I_{\xi }/I_{\chi }\sim
10^{11}$. Thus, for realization of multiphoton SB in graphene one can expect 
$10^{11}$\ times smaller intensities than for SB in atoms \cite{7aa}, \cite%
{7aaa}, \cite{12a,12aa,13}. In the presented work, the influence of
multiphoton effects in SB\ absorption process with an external EM\ wave
field of moderate intensities is considered. Note that the first
nonrelativistic treatment of SB in the Born approximation has been carried
out analytically in the work \cite{12a}, and then this approach has been
extended to the relativistic domain \cite{12aa}.

Let us treat the relativistic quantum theory of graphene nonlinear
interaction with the arbitrary strong EM wave field by microscopic theory
for the 2D Dirac fermions-ions interaction on the base of density matrix. We
consider a wave field exactly and a scattering potential of doped graphene
impurity ions as a perturbation. We consider the interaction when the laser
wave propagates in a perpendicular direction to the graphene plane ($XY$) to
exclude the effect of magnetic field. This traveling wave for electrons in
graphene becomes a homogeneous time-periodic electric field. It is directed
along the $X$ axis with the form (constant phase connected with the position
of the wave pulse maximum with respect to the graphene plane is set zero).
We assume the EM wave to be quazimonochromatic and of the linear
polarization with frequency $\omega $ and amplitude $E_{0}$:%
\begin{equation}
\mathbf{E}(t)=\widehat{\mathbf{x}}E_{0}\cos \omega t.  \label{1_0a}
\end{equation}%
The corresponding wave vector potential $\mathbf{A}(t)$ will be:%
\begin{equation}
\mathbf{A}(t)=-c\int_{0}^{t}\mathbf{E}(t^{\prime })dt^{\prime }=-\widehat{%
\mathbf{x}}\frac{E_{0}}{\omega }\sin \omega t.  \label{1a}
\end{equation}%
To exclude the valence electrons excitations at high Fermi energies in
graphene, we will assume for a EM\ wave actually a terahertz radiation. The
impurity ions are assumed to be at rest and either randomly or nonrandomly
distributed in the doped graphene, the arbitrary form electrostatic
potential field of which is described by the scalar potential:%
\begin{equation}
\varphi (\mathbf{r})=\sum\limits_{i}^{N_{i}}\varphi _{i}(\mathbf{r-R}_{i}).
\label{2a}
\end{equation}%
Here $\varphi _{i}$ is the potential of a single ion placed at the position $%
\mathbf{R}_{i}$, and $N_{i}$\ is the number of impurity ions in the
interaction region.

Let us consider the quantum kinetic equations for a single particle density
matrix for SB process investigation, which can be derived from the second
quantized formalism. As a basis for single particle wave functions we take
the approximate solution of the massless Dirac equation in the strong EM
wave field $\mathbf{A}(t)$, which may be presented in the form: 
\begin{equation}
\Psi _{\mathbf{p}}(\mathbf{r},t)=\exp \left( \frac{i}{\hbar }\mathbf{pr}%
\right) f_{\mathbf{p}}(t).  \label{2}
\end{equation}%
Hear $\mathbf{r=}\left\{ x,y\right\} $ is the 2D-radius vector. For
determination of spinor wave function $f_{\mathbf{p}}$ we will use the
results of the paper \cite{Ishikawa} with the spinor wave function $f_{%
\mathbf{P}}$ determined as follows:%
\begin{equation}
f_{\mathbf{p}}(t)=\frac{1}{\sqrt{2S}}\left( 
\begin{array}{c}
1 \\ 
e^{i\Theta (\mathbf{p}+\frac{e}{c}\mathbf{A}(t))}%
\end{array}%
\right) e^{-i\Omega (\mathbf{p},t)},  \label{3}
\end{equation}%
where the temporal phase $\Omega (\mathbf{p},t)$ (classical action in the
field (\ref{1a})) is defined as: 
\begin{equation}
\Omega (\mathbf{p},t)=\frac{\mathrm{v}_{F}}{\hbar }\int \sqrt{\left( p_{x}+%
\frac{e}{c}A_{x}\right) ^{2}+p_{y}^{2}}dt.  \label{4a}
\end{equation}%
The function $\Theta (\mathbf{p})$ is the polar angle in momentum space and $%
S$ is the quantization area (graphene layer surface area). In terms of these
parameters, the graphene linear dispersion law for quasiparticles
energy-momentum $\mathcal{E}(p)$ defined by the characteristic Fermi
velocity $\mathrm{v}_{F}$, reads: $\mathcal{E}(p)=\pm \mathrm{v}%
_{F}\left\vert \mathbf{p}\right\vert $ $=\pm \mathrm{v}_{F}\sqrt{%
p_{x}^{2}+p_{y}^{2}}$, where the upper sign corresponds to electrons and the
lower sign - to holes. The states (\ref{2}) are normalized by the condition%
\begin{equation}
\int \Psi _{\mathbf{p}^{\prime }}^{+}(\mathbf{r},t)\Psi _{\mathbf{p}}(%
\mathbf{r},t)d\mathbf{r}=\frac{(2\pi \hbar )^{2}}{S}\delta \left( \mathbf{p-p%
}^{\prime }\right) .  \label{5a}
\end{equation}%
The Hamiltonian of the system in the second quantization formalism can be
presented in the form: 
\begin{equation}
\mathcal{H}=\int \widehat{\Psi }^{+}\widehat{H}_{0}\widehat{\Psi }d\mathbf{r+%
}\mathcal{H}_{sb},  \label{8a}
\end{equation}%
where $\widehat{\Psi }$ is the field operator for quasiparticles of the
Fermi-Dirac sea in the graphene, $\widehat{H}_{0}$ is the single-particle 2D
Dirac Hamiltonian in the external field $\mathbf{A}(t)$ (\ref{1a}), and the
interaction Hamiltonian of SB process in the EM wave is%
\begin{equation}
\mathcal{H}_{sb}=\frac{1}{c}\int \widehat{j}\varphi (\mathbf{r})d\mathbf{r,}
\label{8aa}
\end{equation}%
with the current density operator%
\begin{equation}
\widehat{j}=-e\mathrm{v}_{F}g_{s}g_{v}\int \widehat{\Psi }^{+}\widehat{%
\sigma }\widehat{\Psi }d\mathbf{r.}  \label{8aaa}
\end{equation}%
Making Fourier transform of scalar potential 
\begin{eqnarray}
\varphi (\mathbf{r}) &=&\frac{1}{(2\pi )^{2}}\int V(\mathbf{q})e^{-i\mathbf{%
qr}}d\mathbf{q,}  \label{6} \\
V\left( \mathbf{q}\right) &=&\int \sum\limits_{i}^{N_{i}}e\varphi _{i}(%
\mathbf{r-R}_{i})e^{i\mathbf{qr}}d\mathbf{r}  \notag
\end{eqnarray}%
the interaction Hamiltonian can be expressed in the following form 
\begin{equation}
\mathcal{H}_{sb}=-\frac{g_{s}g_{v}\mathrm{v}_{F}}{c(2\pi )^{2}}\int \int 
\widehat{\Psi }^{+}V(\mathbf{q})e^{-i\mathbf{qr}}\widehat{\Psi }d\mathbf{q}d%
\mathbf{r.}  \label{9a}
\end{equation}%
In Eq. (\ref{8aaa}) $\widehat{\sigma }=\left\{ \sigma _{x},\sigma
_{y}\right\} $ - Pauly matrices, $g_{s}$ and $g_{v}$ are the spin and valley
degeneracy factors, respectively.

Let us pass to Furry representation and present the Heisenberg field
operator of the electron in the form of an expansion in the quasistationary
Dirac states (\ref{2}):%
\begin{equation}
\widehat{\Psi }(\mathbf{r},t)=\int d\Phi _{\mathbf{p}}\widehat{a}_{\mathbf{p}%
}\Psi _{\mathbf{p}}(\mathbf{r},t)\mathbf{,}  \label{8}
\end{equation}%
where $d\Phi _{\mathbf{p}}=Sd^{2}\mathbf{p}d\theta \mathbf{/}(2\pi \hbar
)^{2}$. We have excluded the hole operators in Eq. (\ref{8}), since
contribution of electron-holes intermediate states will be negligible for
considered intensities and Fermi energies. The creation and annihilation
operators $\widehat{a}_{\mathbf{P}}^{+}(t)$ and $\widehat{a}_{\mathbf{P}%
}(t), $ associated with positive energy solutions and satisfy the known
anticommutation rules at equal times \cite{Mer2}:%
\begin{equation}
\left\{ \widehat{a}_{\mathbf{p}}^{+}(t)\widehat{a}_{\mathbf{p}^{\prime
}}\left( t^{\prime }\right) \right\} _{t=t^{\prime }}=\frac{(2\pi \hbar )^{2}%
}{S}\delta \left( \mathbf{p}-\mathbf{p}^{\prime }\right) ,  \label{9aaaa}
\end{equation}%
\begin{equation}
\left\{ \widehat{a}_{\mathbf{p}}^{+}(t)\widehat{a}_{\mathbf{p}^{\prime
}}^{+}\left( t^{\prime }\right) \right\} _{t=t^{\prime }}=\left\{ \widehat{a}%
_{\mathbf{p}}(t)\widehat{a}_{\mathbf{p}^{\prime }}\left( t^{\prime }\right)
\right\} _{t=t^{\prime }}=0.  \label{9aa}
\end{equation}%
Taking into account anticommutation rules, Eqs. (\ref{8}), (\ref{2})-(\ref%
{4a}), the second quantized Hamiltonian can be expressed in the form 
\begin{equation}
\mathcal{H}=\mathcal{H}_{0}\mathbf{+}\mathcal{H}_{sb}(t),  \label{9}
\end{equation}%
where the first term is the Hamiltonian of the wave dressed two-dimensional
massless Dirac fermion field: 
\begin{equation}
\mathcal{H}_{0}=\mathrm{v}_{F}\int d\Phi _{\mathbf{p}}P\widehat{a}_{\mathbf{p%
}}^{+}\widehat{a}_{\mathbf{p}},  \label{10}
\end{equation}%
and the second term%
\begin{equation}
\mathcal{H}_{sb}=\int d\Phi _{\mathbf{p}}\int d\Phi _{\mathbf{p}^{\prime
}}C_{\mathbf{p}^{\prime }\mathbf{p}}(t)\widehat{a}_{\mathbf{p}^{\prime }}^{+}%
\widehat{a}_{\mathbf{p}}  \label{11a}
\end{equation}%
is the Hamiltonian of interaction in SB\ process. Here $P$ is the absolute
value of the Dirac particle "quasimomentum" which is defined from kinematic
momentum in EM field $\mathbf{p}$ as:%
\begin{equation}
P=\frac{\omega }{2\pi }\int_{0}^{\frac{2\pi }{\omega }}\sqrt{\left( p_{x}+%
\frac{e}{c}A_{x}(t)\right) ^{2}+p_{y}^{2}}dt.  \label{17a}
\end{equation}%
For the impurity potential of the arbitrary form electrostatic potential $V(%
\mathbf{q})$ from the relation (\ref{8}), (\ref{2})-(\ref{4a}) we have the
following relations for the SB amplitudes:

\begin{equation*}
C_{\mathbf{p}^{\prime }\mathbf{p}}(t)=-\frac{g_{s}g_{v}}{2S}\frac{\mathrm{v}%
_{F}}{c}V\left( \mathbf{q}\right) \left( 1+e^{i\left[ \Theta (\mathbf{p}+%
\frac{e}{c}\mathbf{A}(\tau ))-\Theta (\mathbf{p}^{\prime }+\frac{e}{c}%
\mathbf{A}(t))\right] }\right)
\end{equation*}%
\begin{equation}
\times e^{-\frac{i}{\hbar }\mathrm{v}_{F}\int_{0}^{\tau }\left[ \sqrt{\left(
p_{x}^{\prime }+\frac{e}{c}A_{x}\right) ^{2}+p_{y}^{\prime 2}}-\sqrt{\left(
p_{x}+\frac{e}{c}A_{x}\right) ^{2}+p_{y}^{2}}\right] dt},  \label{7}
\end{equation}%
where $\mathbf{q}=\frac{\mathbf{p}^{\prime }\mathbf{-p}}{\hbar }$ is the
recoil momentum. In accordance to Eq. (\ref{7}) the amplitude $C_{\mathbf{p}%
^{\prime }\mathbf{p}}(t)$ can be expressed in the following form:%
\begin{equation}
C_{\mathbf{p}^{\prime }\mathbf{p}}(t)=\frac{1}{S}e^{-\frac{i}{\hbar }\mathrm{%
v}_{F}\left( P^{\prime }-P\right) t}B(t),  \label{7b}
\end{equation}%
where the time-depended function%
\begin{equation*}
B(t)=-\frac{g_{s}g_{v}\mathrm{v}_{F}}{2c}V\left( \mathbf{q}\right)
\end{equation*}%
\begin{equation*}
\times \left( 1+e^{i\left[ \Theta (\mathbf{p}+\frac{e}{c}\mathbf{A}%
(t))-\Theta (\mathbf{p}^{\prime }+\frac{e}{c}\mathbf{A}(t))\right] }\right)
\end{equation*}%
\begin{equation}
\times e^{-\frac{i}{\hbar }\mathrm{v}_{F}\int_{0}^{t}\left[ \left( \sqrt{%
\left( p_{x}^{\prime }+\frac{e}{c}A_{x}\right) ^{2}+p_{y}^{\prime 2}}%
-P^{\prime }\right) -\left( \sqrt{\left( p_{x}+\frac{e}{c}A_{x}\right)
^{2}+p_{y}^{2}}-P\right) \right] dt^{\prime }}.  \label{17}
\end{equation}%
Making a Fourier transformation of the function $B(t)$\ (\ref{17}) over $t$%
,\ using the known relations%
\begin{equation}
B(t)=\sum\limits_{n=-\infty }^{\infty }e^{-in\omega t}\widetilde{B}_{\mathbf{%
p}^{\prime }\mathbf{p}}^{\left( n\right) },  \label{19}
\end{equation}%
\begin{equation}
\widetilde{B}_{\mathbf{p}^{\prime }\mathbf{p}}^{\left( n\right) }=\frac{%
\omega }{2\pi }\int_{0}^{2\pi /\omega }e^{in\omega t}B(t)dt,  \label{20}
\end{equation}%
we can write the SB amplitude $C_{\mathbf{p}^{\prime }\mathbf{p}}(t)$ (\ref%
{7}) as%
\begin{equation}
C_{\mathbf{p}^{\prime }\mathbf{p}}(t)=\frac{1}{S}\sum\limits_{n=-\infty
}^{\infty }e^{-in\omega t}\widetilde{B}_{\mathbf{p}^{\prime }\mathbf{p}%
}^{\left( n\right) }e^{-\frac{i}{\hbar }\mathrm{v}_{F}\left( P^{\prime
}-P\right) t}.  \label{15}
\end{equation}

To present the microscopic relativistic quantum theory of the multiphoton
inverse-bremsstrahlung absorption of external EM wave radiation in doped
graphene we solve the Liouville-von Neumann equation for the density matrix $%
\widehat{\rho }$:%
\begin{equation}
\frac{\partial \widehat{\rho }}{\partial t}=\frac{i}{\hbar }\left[ \widehat{%
\rho },\mathcal{H}_{0}+\mathcal{H}_{sb}(t)\right]  \label{22a}
\end{equation}%
with the initial condition 
\begin{equation}
\widehat{\rho }(0)=\widehat{\rho }_{g}.  \label{23a}
\end{equation}%
It is assumed that before the interaction with EM wave the system of
graphene quasiparticles was an ideal Fermi gas in equilibrium (thermal and
chemical) with a reservoir. Thus the density matrix $\widehat{\rho }_{g}$ of
grand canonical ensemble is:%
\begin{equation}
\widehat{\rho }_{g}=\exp \left[ \frac{1}{T_{e}}\left( W+\int d\Phi _{\mathbf{%
p}}\left( \mu -\mathrm{v}_{F}P\right) \widehat{a}_{\mathbf{p}}^{+}\widehat{a}%
_{\mathbf{p}}\right) \right] .  \label{24a}
\end{equation}%
In Eq. (\ref{24a}) $T_{e}$\ is the electrons temperature in energy units, $%
\mu $ is the chemical potential, $W$ is the grand potential. The initial
single-particle density matrix in momentum space will be a diagonal, and we
will have the Fermi-Dirac distribution:%
\begin{equation*}
\rho \left( \mathbf{p}_{1},\mathbf{p}_{2},0\right) =Tr\left( \widehat{\rho }%
_{g}\widehat{a}_{\mathbf{p}_{2}}^{+}\widehat{a}_{\mathbf{p}_{1}}\right) =
\end{equation*}%
\begin{equation}
f(P_{1})\frac{(2\pi \hbar )^{2}}{S}\delta \left( \mathbf{p}_{1}-\mathbf{p}%
_{2}\right) ,  \label{25a}
\end{equation}%
where%
\begin{equation}
f(P_{1})=\frac{1}{\exp \left( \frac{\mathrm{v}_{F}P_{1}-\mu }{T_{e}}\right)
+1}.  \label{26a}
\end{equation}%
Within the Born approximation, we consider SB interaction Hamiltonian $%
\mathcal{H}_{sb}(t)$ as a perturbation. So, we expand the density matrix as%
\begin{equation}
\widehat{\rho }=\widehat{\rho }_{g}+\widehat{\rho }_{1},  \label{27a}
\end{equation}%
taking into account the relations%
\begin{equation*}
\left[ \widehat{a}_{\mathbf{p}^{\prime }}^{+}\widehat{a}_{\mathbf{p}},%
\widehat{\rho }_{g}\right] =\left( 1-e^{\frac{\mathrm{v}_{F}}{T_{e}}\left(
P^{\prime }-P\right) }\right) \widehat{\rho }_{g}\widehat{a}_{\mathbf{p}%
^{\prime }}^{+}\widehat{a}_{\mathbf{p}},
\end{equation*}%
\begin{equation*}
\left[ \widehat{\rho }_{g},\mathcal{H}_{0}\right] =0,
\end{equation*}%
for $\widehat{\rho }_{1}$ we obtain:%
\begin{equation*}
\widehat{\rho }_{1}=-\frac{i}{\hbar }\int\limits_{0}^{t}dt^{\prime }\int
d\Phi _{\mathbf{p}}\int d\Phi _{\mathbf{p}^{\prime }}B(t^{\prime })
\end{equation*}%
\begin{equation}
\times e^{\frac{i}{\hbar }\mathrm{v}_{F}\left( t^{\prime }-t\right) \left(
P^{\prime }-P\right) }\left( 1-e^{\frac{\mathrm{v}_{F}}{T_{e}}\left(
P^{\prime }-P\right) }\right) \widehat{\rho }_{g}\widehat{a}_{\mathbf{p}%
^{\prime }}^{+}\widehat{a}_{\mathbf{p}}.  \label{28a}
\end{equation}%
The energy absorption rate of electrons due to inverse bremsstrahlung can be
presented as%
\begin{equation}
\frac{\partial E}{\partial t}=Tr\left( \widehat{\rho }_{1}\frac{\partial 
\mathcal{H}_{sb}(t)}{\partial t}\right) .  \label{29a}
\end{equation}%
It is convenient to represent the rate of inverse bremsstrahlung absorption
via the mean number of absorbed photons per impurity ion, per unit time:%
\begin{equation}
\frac{dN_{abs}}{dt}=\frac{1}{\hbar \omega N_{i}}\frac{\partial E}{\partial t}%
,  \label{30a}
\end{equation}%
where $N_{i}$ is the number of impurity ions in the interaction region.

Taking into account the decomposition relation:%
\begin{equation*}
\left( 1-e^{\frac{\mathrm{v}_{F}}{T_{e}}\left( P_{1}-P_{2}\right) }\right)
Tr\left( \widehat{\rho }_{g}\widehat{a}_{\mathbf{p}_{1}}^{+}\widehat{a}_{%
\mathbf{p}_{2}}\widehat{a}_{\mathbf{p}_{3}}^{+}\widehat{a}_{\mathbf{p}%
_{4}}\right) =
\end{equation*}%
\begin{equation}
\left( 1-e^{\frac{\mathrm{v}_{F}}{T_{e}}\left( P_{1}-P_{2}\right) }\right)
f_{1}\left( 1-f_{2}\right) ,  \label{31a}
\end{equation}%
and making the some calculations using the relations (\ref{28a})-(\ref{31a}%
), (\ref{11a}), (\ref{7b}), (\ref{17}) for large $t$\ we obtain:%
\begin{equation}
\frac{dN_{abs}}{dt}=\sum\limits_{n=1}^{\infty }\frac{dN_{abs}\left( n\right) 
}{dt},  \label{31aa}
\end{equation}%
where the partial absorption rates have the following forms%
\begin{equation*}
\frac{dN_{abs}\left( n\right) }{dt}=\frac{4\pi g_{s}g_{v}n}{\hbar N_{i}S^{2}}%
\frac{\mathrm{v}_{F}}{c}\int \int d\Phi _{\mathbf{p}}d\Phi _{\mathbf{p}%
^{\prime }}\left\vert V\left( \mathbf{q}\right) \right\vert ^{2}\left\vert 
\widetilde{M}_{\mathbf{p}^{\prime }\mathbf{p}}^{\left( n\right) }\right\vert
^{2}
\end{equation*}%
\begin{equation*}
\times \delta \left( \mathrm{v}_{F}P^{\prime }-\mathrm{v}_{F}P+n\hbar \omega
\right) \left( 1-e^{\frac{\mathrm{v}_{F}}{T_{e}}\left( P^{\prime }-P\right)
}\right)
\end{equation*}%
\begin{equation}
\times f\left( P^{\prime }\right) \left( 1-f\left( P\right) \right) ,
\label{32a}
\end{equation}%
where 
\begin{equation*}
\left\vert \widetilde{M}_{\mathbf{p}^{\prime }\mathbf{p}}^{\left( n\right)
}\right\vert ^{2}=\left\vert \int_{0}^{T}d\left( \frac{t}{T}\right) \left(
1+e^{i\left[ \Theta (\mathbf{p}+\frac{e}{c}\mathbf{A}(t))-\Theta (\mathbf{p}%
^{\prime }+\frac{e}{c}\mathbf{A}(t))\right] }\right) \exp (in\omega t)\right.
\end{equation*}%
\begin{equation}
\times \left. e^{-\frac{i}{\hbar }\mathrm{v}_{F}\int_{0}^{t}\left[ \left( 
\sqrt{\left( p_{x}^{\prime }+\frac{e}{c}A_{x}\right) ^{2}+p_{y}^{\prime 2}}%
-P^{\prime }\right) -\left( \sqrt{\left( p_{x}+\frac{e}{c}A_{x}\right)
^{2}+p_{y}^{2}}-P\right) \right] dt^{\prime }}\right\vert ^{2}.  \label{33a}
\end{equation}%
In Eq. (\ref{32a}) $\delta \left( x\right) $ is the Dirac function $\delta $
that expresses the energy conservation law of the SB process. The obtained
formula for the absorption rate is true for grand canonical ensemble and
positive only. With the help of (\ref{33a}) one can investigate the
nonlinear inverse-bremsstrahlung absorption rate for degenerate quantum
plasma state - the graphene electrons with the distribution function given
by the Eq. (\ref{26a}).

\section{Numerical results for SB absorption coefficient for the screened
Coulomb potential of impurity ions in graphene}

Now we utilize Eq. (\ref{33a}) in order to obtain the inverse-bremsstrahlung
absorption coefficient in particular case of SB process on a screened
Coulomb potential of impurity ions in graphene \cite{BornEl2}, \cite%
{Katsnelson}, \cite{Adam2007,Ando1982,Saito,Wang2007}. In general, one
should arise from the Linhard theory of screening, which is applicable for
quantum plasmas \cite{46}. For our calculations the Thomas-Fermi
approximation is valid, and the potential of a single ion $\varphi _{i}$
varies slowly on the scale of Fermi wavelength $\hbar /\mathrm{p}_{F}$,
where $\mathrm{p}_{F}$\ is the Fermi momentum. In accordance with \cite%
{BornEl2}, \cite{PhysRev2015}, for randomly distributed charged impurity
ions we have the Fourier transform of potential (\ref{6}):%
\begin{equation}
\left\vert V\left( \mathbf{q}\right) \right\vert ^{2}=N_{i}\frac{4\pi
^{2}e^{4}}{\widetilde{\kappa }^{2}q^{2}\epsilon ^{2}\left( q\right) },
\label{29}
\end{equation}%
where $\epsilon \left( q\right) $ ($q=\left\vert \mathbf{q}\right\vert $) is
the 2D static dielectric (screening) function in random phase approximation
(RPA) appropriate for graphene \cite{Wang2007}, given by the formula%
\begin{equation*}
\epsilon \left( q\right) =1+\frac{q_{s}}{q}
\end{equation*}%
\begin{equation}
\times \left\{ 
\begin{array}{c}
1-\frac{\pi q}{8k_{F}},\quad q\leq 2k_{F} \\ 
1-\frac{\sqrt{q^{2}-4k_{F}^{2}}}{2q}-\frac{q\sin ^{-1}2k_{F}/q}{4k_{F}}%
,\quad q>2k_{F}%
\end{array}%
\right. .  \label{26}
\end{equation}%
Here $k_{F}=\varepsilon _{F}/\hbar \mathrm{v}_{F}$ is 2D Fermi wave vector, $%
q_{s}=4e^{2}k_{F}/\left( \hbar \widetilde{\kappa }\mathrm{v}_{F}\right) $ is
the effective graphene 2D Thomas-Fermi wave vector, and $\widetilde{\kappa }$
$\mathbf{=}$ $\kappa \left( 1+\pi r_{s}/2\right) $ is the\ effective
dielectric constant of a substrate. The ratio of the potential to the
kinetic energy in an interacting quantum Coulomb system is measured by the
dimensionless Wigner-Seitz radius $r_{s}=e^{2}/\kappa \hbar \mathrm{v}_{F}$,
where $\kappa $ is the background lattice dielectric constant of the system, 
$e^{2}/\hbar \mathrm{v}_{F}\simeq 2.18$ is \textquotedblleft effective
fine-structure constant\textquotedblright\ in graphene (in the vacuum). 
\begin{figure}[tbp]
\centering{\includegraphics[width=.51\textwidth]{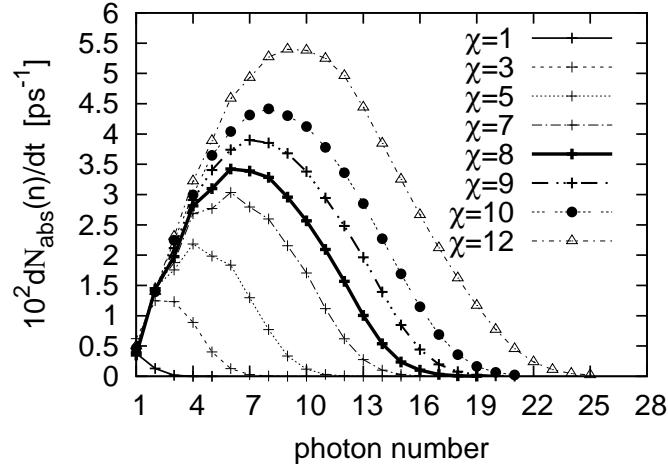}}
\caption{Envelope of partial rate $dN_{abs}\left( n\right) /dt$ of inverse
bremsstrahlung absorption vs the mean number of absorbed photons by per ion,
per unit time (in \textrm{ps}$^{-1}$) for linear polarization of EM wave in
doped graphene is shown for various wave intensities at $\protect\varepsilon %
\equiv \hbar \protect\omega =0.01$ \textrm{eV, }$T_{e}=0.1\protect%
\varepsilon _{F}$, and $\protect\varepsilon _{F}\simeq \protect\mu =20%
\protect\varepsilon $.}
\label{11}
\end{figure}
\begin{figure}[tbp]
\centering{\includegraphics[width=.51\textwidth]{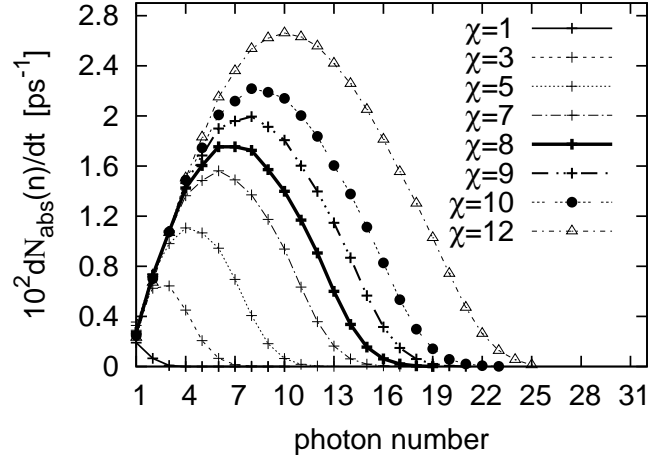}}
\caption{Same as Fig. 1 but for photon energy $\protect\varepsilon =0.005$\ $%
\mathrm{eV.}$}
\label{222}
\end{figure}
Taking into account Eqs. (\ref{33a})-(\ref{26}), (\ref{26a}), and
integrating in Eq. (\ref{32a}) over $P^{\prime }$, we obtain the following
relation for the partial absorption rates $dN_{abs}\left( n\right) /dt$ of
SB process:%
\begin{equation*}
\frac{dN_{abs}\left( n\right) }{dt}=\frac{g_{s}g_{v}n\mathrm{v}_{F}^{2}}{\pi
\hbar c}\left( \frac{r_{s}}{2+\pi r_{s}}\right) ^{2}\int_{\mathrm{v}%
_{F}P+n\hbar \omega }dP\int \int d\theta d\theta ^{\prime }\frac{\left\vert 
\mathbf{P}\right\vert \left\vert \mathbf{P}^{\prime }\right\vert \left\vert 
\widetilde{M}_{\mathbf{p}^{\prime }\mathbf{p}}^{\left( n\right) }\right\vert
^{2}}{\left( \hbar q\right) ^{2}\epsilon ^{2}\left( q\right) }
\end{equation*}%
\begin{equation}
\times \left( 1-e^{-\frac{n\hbar \omega }{T_{e}}}\right) f\left( \mathrm{v}%
_{F}P-n\hbar \omega \right) \left( 1-f\left( \mathrm{v}_{F}P\right) \right) .
\label{exac}
\end{equation}%
At the consideration of numerical results it is convenient to represent the
differential cross-sections of SB on the charged impurities in the form of
dimensionless quantities. For the dimensionless rates $TdN_{abs}\left(
n\right) /dt$ in the field of linearly polarized EM\ wave with the
dimensionless vector potential $\overline{\mathbf{A}}(t)=-\widehat{\mathbf{x}%
}{\chi }\sin (2\pi \tau )$ we have $\mathrm{:}$%
\begin{equation*}
T\frac{dN_{abs}\left( n\right) }{dt}=\frac{2g_{s}g_{v}n}{300}\left( \frac{%
r_{s}}{2+\pi r_{s}}\right) ^{2}\int_{\overline{P}+n}d\overline{P}\int \int
d\theta _{\mathbf{p}}d\theta _{\mathbf{p}^{\prime }}\frac{\overline{P}\left( 
\overline{P}-n\right) \left\vert \overline{M}_{\mathbf{p}^{\prime }\mathbf{p}%
}^{\left( n\right) }\right\vert ^{2}}{\left\vert \overline{\mathbf{q}}%
\right\vert ^{2}\epsilon ^{2}\left( \left\vert \overline{\mathbf{q}}%
\right\vert \right) }
\end{equation*}%
\begin{equation}
\times \left( 1-e^{-\frac{n}{\overline{T_{e}}}}\right) f\left( \overline{P}%
-n\right) \left( 1-f\left( \overline{P}\right) \right) ,  \label{exact2}
\end{equation}%
where 
\begin{equation*}
\left\vert \overline{M}_{\mathbf{p}^{\prime }\mathbf{p}}^{\left( n\right)
}\right\vert ^{2}=\left\vert \int_{0}^{1}d\tau \left( 1+e^{i\left[ \Theta (%
\overline{\mathbf{p}^{\prime }}-\widehat{\mathbf{x}}{\chi }\sin (2\pi \tau
))-\Theta (\overline{\mathbf{p}}-\widehat{\mathbf{x}}{\chi }\sin (2\pi \tau
))\right] }\right) \right.
\end{equation*}%
\begin{equation*}
\times \exp \left\{ i2\pi n\tau -2\pi i\int_{0}^{\tau }\left[ \left( \sqrt{%
\left( \overline{p}_{x}^{\prime }-{\chi }\sin (2\pi \tau ^{\prime })\right)
^{2}+\overline{p}_{y}^{\prime 2}}-\overline{P}^{\prime }\right) \right.
\right.
\end{equation*}%
\begin{equation}
\left. \left. \left. -\left( \sqrt{\left( \overline{p}_{x}-{\chi }\sin (2\pi
\tau ^{\prime })\right) ^{2}+\overline{p}_{y}^{2}}-\overline{P}\right) %
\right] d\tau ^{\prime }\right\} \right\vert ^{2}.  \label{333}
\end{equation}%
In Eq. (\ref{exact2}) the dimensionless momentum, energy, time, and
relativistic invariant intensity parameter of EM wave introduced as follows:%
\begin{equation*}
\overline{p}_{x,y}=\frac{\mathrm{v}_{F}}{\hbar \omega }p_{x,y},\overline{%
\mathcal{E}}=\frac{\mathcal{E}}{\hbar \omega },\overline{\mu }=\frac{\mu }{%
\hbar \omega },\overline{T}_{e}=\frac{T_{e}}{\hbar \omega },\overline{P}=%
\frac{\mathrm{v}_{F}}{\hbar \omega }P,
\end{equation*}%
\begin{equation}
\overline{k}_{F}=\frac{\mathcal{E}_{F}}{\hbar \omega },d\tau =\frac{dt}{T},{%
\chi }=\frac{e\mathrm{v}_{F}}{\hbar \omega ^{2}}E_{0}.  \label{23}
\end{equation}

The analytic integration over scattering angles $d\theta _{\mathbf{p}%
},d\theta _{\mathbf{p}^{\prime }}$ and momentum is impossible, so we make
numerical integrations. For numerical analysis of SB\ cross sections in
graphene we assume Fermi energy $\varepsilon _{F}\simeq \mu =20\hbar \omega $
($\varepsilon _{F}\gg n\hbar \omega $), electrons temperature $%
T_{e}=0.1\varepsilon _{F}$, coherent EM linearly polarized radiation,
dielectric environment constant $\kappa =2.5$ for an impurity strength in
the presence of the $\mathrm{SiO}_{2}$ substrate \cite{Wang2007},
Wigner-Seitz radius $r_{s}=0.87592$.

In the Fig. 1 and Fig. 2 the envelope of partial rate of
inverse-bremsstrahlung absorption in graphene is shown for various wave
intensities with energy of photons $\varepsilon \equiv \hbar \omega =0.01$ 
\textrm{eV }($\lambda =1.24\times 10^{-2}\mathrm{cm}$) and $\varepsilon
=0.005$ \textrm{eV }($\lambda =2.48\times 10^{-2}\mathrm{cm}$),
respectively. As seen from these figures, the multiphoton effects become
essential with the increase of the wave intensity.

To show the dependence of the inverse bremsstrahlung absorption rate on the
laser radiation intensity, the total SB\ rate (\ref{31aa}) via the mean
number of absorbed photons by per of impurity ion, per unit time in doped
graphene versus the parameter ${\chi }$ for various photon energies is shown
in Fig. 3. To compare with the linear theory \cite{12a}, in Fig. 4 we plot
scaled absorption rate ${\chi }^{-2}dN_{abs}/dt$ versus ${\chi }$. As it is
presented by the figure, for the large values of ${\chi }$ the SB rate
exhibits a slightly falling dependence on the wave intensity. To all other,
in the scope of linear theory the scaled absorption rate does not depend on
the wave intensity, while for the large values $\chi $ it is suppressed with
the increase of the wave intensity for the multiphoton SB process.

As was expected and as seen from these figures with the increasing of laser
intensity the multiphoton effect becomes dominant compared to the one-photon
scattering in linear theory (\cite{12a}). For THz photons, the multiphoton
interaction regime in graphene can be achieved already at the intensities $%
I_{\chi }\sim 10^{3}$\ $Wcm^{-2}$. Thus, for these intensities multiphoton
SB process opens new channels for the wave absorption, and we can expect
strong deviation of absorbance of a single layer doped graphene from linear
one, which for frequencies smaller than Fermi energy is zero \cite{trans}.
The latter opens up possibility for manipulating of electronic transport
properties of the doped graphene by coherent radiation field. 
\begin{figure}[tbp]
\centering{\includegraphics[width=.51\textwidth]{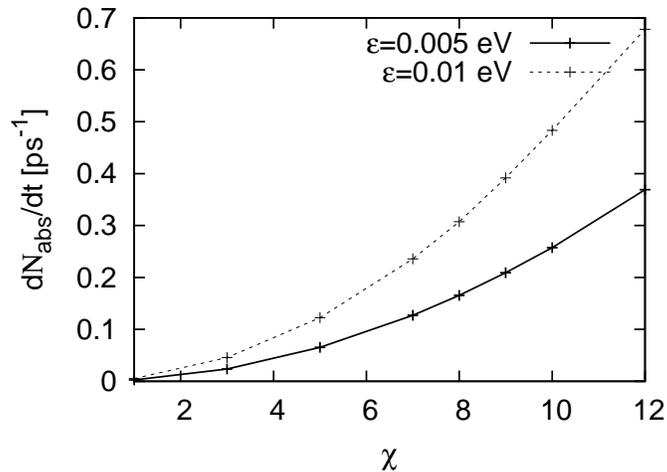}}
\caption{{}Total rate of inverse bremsstrahlung in doped graphene vs the
dimensionless parameter $\protect\chi $\ for setup of Fig. 1 at various
photon energies $\protect\varepsilon $.}
\label{33}
\end{figure}
\begin{figure}[tbp]
\centering{\includegraphics[width=.51\textwidth]{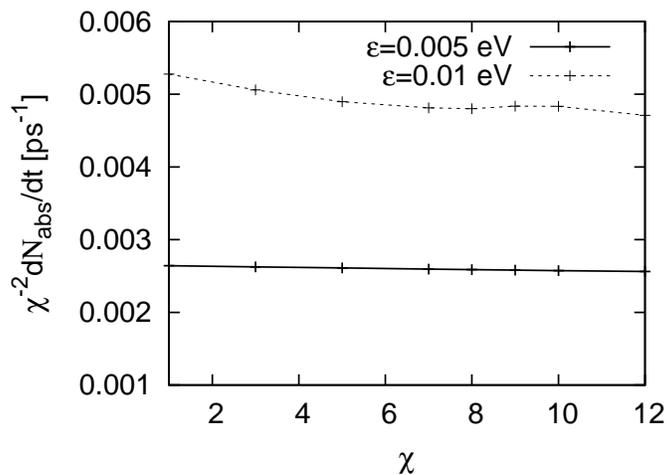}}
\caption{Total rates of the inverse bremsstrahlung absorption scaled to $%
\protect\chi ^{2}$\ vs the parameter $\protect\chi $.}
\label{44}
\end{figure}

\section{Conclusion}

We have presented the microscopic relativistic quantum theory of multiphoton
SB absorption in doped graphene in the presence of the coherent EM radiation
of arbitrary intensity and frequency (actually terahertz radiation to
exclude the valence electrons excitations at high Fermi energies). The
Liouville-von Neumann equation for the density matrix has been solved
analytically considering an external wave-field exactly, and the charged
impurity ions arbitrary electrostatic potential in the Born approximation.
These solutions for SB at the linear polarization of EM wave are used for
derivation of a relatively compact formula for the nonlinear
inverse-bremsstrahlung absorption rate when 2D Dirac fermions are
represented by the grand canonical ensemble. The obtained relativistic
analytical formulas have been analyzed numerically for screened Coulomb
potential. The concluded results show that SB rate in graphene in the
presence of strong terahertz radiation field being practically independent
of the plasma Fermi energy, has an essentially nonlinear dependence on the
increase of the wave intensity, and the multiphoton absorption/emission
processes play a significant role already at moderate laser intensities. It
is shown that one can achieve the efficient absorption coefficient by this
mechanism for these intensities.

\begin{acknowledgments}
The authors are deeply grateful to the Prof. Hamlet K. Avetissian and Dr.
G.F. Mkrtchian for permanent discussions during the work on the present
paper, for valuable comments and recommendations. This work was supported by
the State Committee of Science MES RA together with the Fundamental Research
Foundation of the Republic of Belarus, in the frame of the research project
No. AB16-19.
\end{acknowledgments}

\end{document}